\documentclass[11pt]{article}
\usepackage[T1]{fontenc}
\usepackage[latin2]{inputenc}
\usepackage{amsmath}
\usepackage{amssymb}
\usepackage{times}
\usepackage[normalem]{ulem}
\def\tr{{\rm tr}}
\def\d{{\rm d}}
\def\om{\omega}
\def\a{\alpha}
\def\half{{\scriptstyle\frac{1}{2}}}
\def\ket#1{|#1\rangle}
\def\bra#1{\langle#1|}

\begin{document}
\title{Non-Markovian Open Quantum Systems: Lorentzian from Markovian}
\author{Lajos Di\'osi\\Research Institute for Particle and Nuclear Physics\\
H-1525 Budapest 114, POB 49, Hungary}
\date{\today}
\maketitle
\begin{abstract}As a general mission, reduced dynamics and master equations are advocated as alternative method
 and philosophy instead of Green functions, Kubo theory and the like.
 A smart reduction of the Lorentzian open system to the \sout{Ohmic} Markovian one (Imamo{\~g}lu, 1994) is presented 
in simple terms.

The original paper appeared in Hungarian, in \emph{New Results in Quantum Optics and Electronics}, eds.: Zsuzsanna Heiner and K. Osvay (University of Szeged, Szeged, 2006) p147.
\end{abstract}

\section{Introduction: open$=$reduced}
Numerous physical phenomena can be modeled as open dynamical systems.
Typically, the very system S under investigation is interacting with a reservoir 
R and \emph{this} makes S dynamically open.  Its dynamics changes, becomes
in general irreversible, dissipative, and non-Markovian. The total composite
system S+R can be treated as a closed dynamical system. Its state $\rho_{SR}$ 
evolves reversibly. In interaction picture, the unitary evolution of $\rho_{SR}$ 
is generated by the interaction Hamiltonian $H_{SR}$:
\begin{equation}
\frac{\d}{\d t}\rho_{SR}(t)=-i\left[H_{SR}(t)~,\rho_{SR}(t)\right].
\end{equation}
From it, we can derive the so-called reduced state of the system S: 
\begin{equation}
\rho_S(t)=\tr_R\rho_{SR}(t),
\end{equation}
and its reduced dynamics as well. If the reservoir's self-dynamics and its coupling to
the system are of simple structure  then the reduced dynamics will be well calculable.
Suppose the system and the reservoir are initially at time  $t=0$ uncorrelated,
so the system's reduced dynamics can be described by a homogeneous non-Markovian 
master equation \cite{Nakajima}:
\begin{equation}
\frac{\d}{\d t}\rho_S(t)=\int_0^t \d t'{\cal K}(t-t')\rho_S(t'),
\end{equation}
which is in principle solvable if we know the initial state  $\rho_S(0)$ of the system.
However, the closed analytic form of the memory kernel ${\cal K}(t-t')$ can be found
in special cases only. If the self-dynamics of the system is slow compared to the
characteristic time of the said memory kernel then the so-called Markovian approximation
turns the master equation into the much simpler form:
\begin{equation}
\frac{\d}{\d t}\rho_S(t)={\cal L}\rho_S(t).
\end{equation}
That is a remarkable limit case. It was applied for classical Brownian motion, as well as
to quantum physics, for spontaneous atomic emission, nuclear spin-relaxation, transport
phenomena, the whole of quantum optics, from quantum electronics to molecules,
actually for any quantum system subject to thermal dissipation (or to decoherence, in novel 
approach) \cite{Weiss}. 

\section{Do we need master equations?}
The number and literature of phenomena that we can model by the system-reservoir 
paradigm is unconceivable. A tendency is, however, seen clearly. Most applications 
avoid the explicite use of the reduced dynamics and of the master equations (3,4).
Rather they concentrate directly on the phenomenon to be described.  Typical
examples are Kubo's theory of response functions, the input-output formalism in
quantum  optics, numerous applications of scattering theory and Green function
techniques. These are presumably the optimum methods for the concrete phenomena. 
Yet, the common divisor of all techniques must be the reduced dynamics, i.e., the
master equation that evolves the state $\rho_S(t)$ of the open system. 
If we point out this explicitely we can deepen our understanding the above
techniques and we can remember the common divisor. We help transitions between
different applications. We help discovery.

Consider, e.g., the problem of electric conductance in metals. The Markovian
master equation (4) of the classical phase space distribution of the electron 
is the classical Boltzmann-Fokker-Planck equation. If we know its solution in
external electric field we can derive Ohm's Law and the conductance. On the
other hand, the widely used modern treatment is based on the Kubo equation.
This establishes a relationship between the external field and the induced
current, and it derives conductance directly from the equilibrium
correlation function of the current. It circumvents the explicite determination of the
reduced dynamics and it only concentrates on the phenomenon of electric 
conductance in question. Not too often, however, we see the conscious parallel
mention of the Boltzmann and Kubo equations \cite{Mahan}. 

The Markovian approximation of the reduced dynamics and its master equation (4)
are well known concerning the mathematical structure \cite{Lindblad}, the
corresponding physics \cite{Weiss}, as well as concerning the methods of
solutions. The clearest conceptual basis of the Markovian approximation must be  
the master equation even if in practice some other serviceable techniques are applied.
When, however, the Markovian approximation is not justified then we cannot
make such a categoric declaration about the credit of master equations. The
structure of non-Markovian master equations (3) has not been exhaustively
studied. Most non-Markovian master equations are difficult and practically 
unsolvable. Thus we cannot say that the explicite use of reduced dynamics 
would always be more advantageous compared to the current techniques
of non-Markovian phenomena.  In this difficult topics, the goal of this talk will be 
limited to flash a genuine smart method for the master equation treatment
of some special non-Markovian reduced dynamics.

\section{Non-Markovian}
Let the reservoir R consist of harmonic modes with spectral density $D(\om)$, with 
emission and absorption operators $b_\om,b_\om^\dagger$, respectively. Let R
couple linearly to a certain quantity $V$ of the system S:
\begin{equation}
H_{SR}=\int 
g_\om\left(V^\dagger b_\om + V b_\om^\dagger\right)D(\om)\frac{\d\om}{2\pi}.
\end{equation}
Assume the uncorrelated initial state:
\begin{equation}
\rho_{SR}(0)=\rho_S(0)\rho_R(0),
\end{equation}
where the system's initial state is arbitrary while the reservoir's is the canonical 
Gibbs equilibrium state. For the sake of simplicity, we follow up the case of zero
temperature. We are looking for the reduced dynamics, i.e., the exact master
equation,  to evolve  $\rho_S(t)$. Consider the field quantity 
$F=\int g_\om b_\om D(\om)\d\om/2\pi$ of the reservoir, that couples linearly
to the system's quantity $V$. In interaction picture this field reads
\begin{equation}
F(t)=\int g_\om b_\om e^{-i\om t}D(\om)\frac{\d\om}{2\pi}.
\end{equation}
Its expectation value vanishes in the reservoir's initial state. The following
correlation function will, however, play a role: 
\begin{equation}
\a(\tau-s)=tr_R \left( F(\tau)F^\dagger(s) \rho_R(0) \right).
\end{equation}
It is known that \emph{the reduced dynamics in interaction picture will depend
but un this correlation function}! We can add that this correlation function is the
Fourier transform of $g_\om^2D(\om)$, i.e., of the reservoir spectral density weighted 
by the coupling strengths.  

In general, the above correlation function is complicated and the explicite form it
determines the requested dynamics (3) requires approximate methods. An exception
is when the system S is simple, e.g., a two state system or being itself, too, a harmonic oscillator.    
In such cases the master equation (3) takes tractable analytic forms. The most
relevant special case is the Markovian one, of course.

\section{Markovian}
Often the spectrum of the reservoir R is broad, smooth and flat compared to the relevant
transition frequencies of the system S. In other words: the reservoir correlation function (8) 
is characterized by a memory time and this is much shorter than the characteristic time
scales of the system's self-dynamics. That is the Markovian limit. In this case therefore
the spectrum (weighted by coupling strengths) will be considered flat and the correlation
function of the field will be considered delta-function:  
\begin{equation}
g_\om^2 D(\om) = f^2={\rm const},~~~~~~~~\a(\tau-s)=f^2\delta(\tau-s).
\end{equation}
\sout{Note that the spectral \emph{energy density} $\om D(\om)$ can alternatively be used,  
the form $g_\om^2 \om D(\om) = f^2\om$, linear in frequency, is called the Ohmic 
spectrum with a reference to the pioneering theory of electric conductance.}
It is known that in the Markovian limit (9) the master equation (4) becomes: 
\begin{equation}
\frac{\d}{\d t}\rho_S=f^2
\left(V\rho_S V^\dagger-\half V^\dagger V\rho_S-\half\rho_S V^\dagger V\right).
\end{equation}
Many times, the solution of such a simple equation is analytically possible.
If not, then a peculiar Monte-Carlo method applies \cite{Schack}, particularly suitable for
solutions of Markovian master equations.  

Let us consider an elementary Markovian example which we need later. Let the system S
itself be an $\om_0$-frequency harmonic oscillator and let its emission operator   
$V=a$ couple to a Markovian \sout{(Ohmic)} reservoir. If the coupling constant is  $f$
then we get the master equation of the damped oscillator which in interaction picture
reads:
\begin{equation}
\frac{\d}{\d t}\rho_A=f^2
\left(a\rho_S a^\dagger-\half a^\dagger a\rho_S-\half\rho_S a^\dagger a\right).
\end{equation}
The ground state $\rho_A=\ket{0;A}\bra{0;A}$ is stationary. In it the expectation value
of the emission and absorption operator vanish. We can determine the ground state 
correlation function of  $a(t)$ and  $a^\dagger(t)$ in interaction picture:
\begin{equation}
\exp[-i\om_0(\tau-s)-\half f^2 \vert\tau-s\vert].
\end{equation}

\section{Reducing Non-Markovian to Markovian}
Now, we tear ourselves away from the Markovian limit and consider a definitely
non-Markovian case which keeps, nonetheless, an intrinsic relationship with the
Markovian ones. Let the spectrum be Lorentzian:
\begin{equation}
g_\om^2 D(\om)= g^2\frac{\gamma}{(\om-\om_0)^2+(\gamma/2)^2}~.
\end{equation}
The corresponding correlation function is exponentially damped:
\begin{equation}
\a(\tau-s)= g^2 \exp[-i\om_0(\tau-s)-\half\gamma\vert\tau-s\vert].
\end{equation}
It can be shown that the reduced dynamics of the system in the Lorentzian
reservoir is exactly identical with the reduced dynamics of a fictitious system
in a Markovian reservoir \cite{Imamoglu}.

Let us first consider a fictitious reservoir consisting of a single $\om_0$-frequency 
oscillator (ancilla). Let us couple the system S to this ancilla instead of the Lorentzian
reservoir:
 \begin{equation}
H_{SA}=g\left(V^\dagger a + V a^\dagger\right).
\end{equation}
This, too, is a system-reservoir interaction just the reservoir consists of a
single oscillator. Otherwise we can do everything like for the case of 
S+R, just we use the notation S+A this time. The coupled field (7) is simply
$F=ga$. In interaction picture we calculate the correlation function (8) of the
field, it is trivial:
\begin{equation}
\bra{0;A} F(\tau)F^\dagger(s)\ket{0;A}=g^2 e^{-i\om_0(\tau-s)}.
\end{equation}
If his correlation function \emph{were} identical with the Lorentzian damped
correlation function (14) then, according to our previous consideration, also the 
reduced dynamics $\tr_A\rho_{SA}(t)$ {\it would} be identical with the reduced dynamics
emerging from a Lorentzian reservoir. But the correlation function of the ancilla A is
not Lorentzian. Yet it can easily be made so! Indeed, we saw that if we place the ancilla-oscillator 
into a\sout{n Ohmic} Markovian reservoir (and choose $f^2=\gamma$) then the ancilla's correlation function
will obtain exactly the desired exponential damping factor (12). It is thus clear that the
influence of the Lorentzian reservoir can exactly be replaced by the influence of an
ancilla-oscillator which itself is damped by a Markovian \sout{(Ohmic)} reservoir.

Let us summarize the method. Consider the composite system $S+A$ with an uncorrelated 
initial state:
\begin{equation}
\rho_S(0)\ket{0;A}\bra{0;A}.
\end{equation}
Assume an interaction $H_{SA}$ (15). Let the ancilla-oscillator A be damped by a\sout{n
Ohmic} Markovian reservoir. Hence the composite state $\rho_{SA}(t)$ evolves according to the 
following Markovian master equation:
\begin{equation}
\frac{\d}{\d t}\rho_{SA}
=-ig\left[ V^\dagger a + V a^\dagger~,~\rho_{SA} \right]
+\gamma
\left(a\rho_{SA}a^\dagger-
\half a^\dagger a\rho_{SA}-\half\rho_{SA}a^\dagger a\right),
\end{equation}
where $a$ and $V$ are time-dependent operators in interaction picture.
If we have solved the above initial value problem then we trace over the
ancilla-oscillator A and this way we obtain the system's current state:
\begin{equation}
\rho_S(t)=\tr_A\rho_{SA}(t).
\end{equation}
Using this method, the non-Markovian influence of a Lorentzian reservoir has
been reduced to the influence of a Markovian \sout{(Ohmic)} reservoir - with the
insertion of a single ancilla-oscillator.

The relevance of Lorentzian reservoir may perhaps not go much beyond the 
high-frequency (Drude-) regularization of the \sout{Ohmic} Markovian one. On the other hand, 
the above method is not restricted to the Lorentzian reservoir, it has been
suggested for the so-called fotonic band gap materials as well since,
surprisingly, the method of ancilla-oscillator applies, with minor modification, 
to an upside-down Lorentzian spectrum as well, i.e., to a Lorentzian forbidden
band in the \sout{Ohmic} flat spectrum. 

\section{Closing remarks}
It was my studying basic quantum structures and phenomena - like the
nowadays popular decoherence or the even more fashionable quantum information 
-  which lead me to open quantum systems and master equations.
I admittedly miss my overview of the corresponding vast literature. Neither
had I felt necessary to disperse the audience's attention by flashing the
investigations where I tried to contribute myself. Rather I thought  
to pick up and present a unique smart non-Markovian method. Also the
cited references are selective - yet suitable to direct the interested toward
the relevant discussions of a few issues on non-Markovian open system
dynamics. 

Acknowledgment: Professor Tam{\'a}s Geszti brought the monograph \cite{Mahan} 
to my attention. My work was supported by the Hungarian Scientific Research
Fund, Grant  T49384 and T75129.

\end{document}